\documentclass[%
 preprint,
 amsmath,amssymb,
 aps,
]{revtex4-2}

\usepackage{graphicx}
\usepackage{dcolumn}
\usepackage{bm}
\usepackage{color}

\begin{document}

\title{Epitaxial binding and strain effects of monolayer stanene on the Al$_{2}$O$_{3}$(0001) surface}

\author{Stephen Eltinge}
 \email{stephen.eltinge@yale.edu}
\affiliation{Department of Physics, Yale University, New Haven, Connecticut 06511, USA
} 
\author{Sohrab Ismail-Beigi}
\affiliation{Department of Applied Physics and Department of Physics, Yale University, New Haven, Connecticut 06511, USA
}

\date{\today}

\begin{abstract}
Stanene, the two-dimensional monolayer form of tin, has been predicted to be a 
2D topological insulator due to its large spin--orbit interaction. However, a 
clear experimental demonstration of stanene's topologically nontrivial properties has eluded 
observation, in part because of the difficulty of choosing a substrate on which 
stanene will remain topologically nontrivial. In this paper, we present 
first-principles density functional theory (DFT) calculations of epitaxial 
monolayer stanene grown on the (0001) surface of alumina, Al$_{2}$O$_{3}$,
as well as free-standing decorated stanene under strain.
By describing the energetics and nature of how  monolayer stanene binds to
alumina, we show a strong energetic drive for the monolayer to be coherently
strained and epitaxial to the substrate. By analyzing the electronic structure
of strained stanene, we find it to be a quantum spin Hall insulator on
Al$_{2}$O$_{3}$. We also describe the effect of \emph{in situ} fluorine decoration on the bound stanene monolayer, including on its potential for mechanical exfoliation.

\end{abstract}

\maketitle


\section{Introduction}

Two-dimensional topological insulators (2DTIs) have received  attention in 
recent years due to their potential for hosting robust  symmetry-protected 
current-carrying edge states \cite{molle_buckled_2017}. The  buckled hexagonal 
monolayer form of tin, known as stanene, is of particular interest 
\cite{lyu_stanene:_2019, sahoo_perspective_2019} since its  band gap 
($\sim0.1$~eV) is large enough for room-temperature applications 
\cite{ma_intriguing_2012, xu_large-gap_2013}. Stanene's band gap can be 
further enhanced by functionalization, in particular with halogen atoms. 
Proposed uses of stanene include spintronic nanoribbon devices 
\cite{zheng_spin-current_2018, marin_tunnel-field-effect_2018}, tunable  
field-effect transistors \cite{vandenberghe_imperfect_2017}, a surface for  
adsorption of molecules including CH$_{2}$O, CH$_{4}$, CO, NO, N$_{2}$O, and  
NH$_{3}$ \cite{abbasi_modulation_2019, abbasi_theoretical_2019}, and  
the possibility of room-temperature demonstration of  the quantum spin Hall
effect \cite{xu_large-gap_2013, zhang_room_2017} and quantum anomalous Hall
effect \cite{wu_prediction_2014, matusalem_quantum_2016, li_stanene_2019}.

However, the electronic structure of epitaxial stanene is sensitive to  both 
strain and surface interactions, so choosing an appropriate substrate is  vital 
\cite{xu_large-gap_2015}. Stanene is metallic on many  substrates, including 
Ag(111) \cite{yuhara_large_2018,  ogikubo_continuous_2020}, Au(111) 
\cite{liu_realization_2019,  pang_epitaxial_2020}, Sb(111) 
\cite{gou_strain-induced_2017}, and  Bi$_{2}$Te$_{3}$(111) 
\cite{zhu_epitaxial_2015, li_anisotropic_2020,  li_-plane_2020}. Ultraflat 
stanene grown on Cu(111) shows evidence of nontrivial edge states but is 
metallic overall \cite{deng_epitaxial_2018},  while buckled stanene on 
PbTe(111) is gapped but is topologically trivial (\emph{i.e.}, non-topological)
due  to in-plane compressive 
strain \cite{zang_realizing_2018}. InSb(111) is a promising substrate for 
globally gapped  topologically nontrivial stanene, though reported results remain somewhat 
inconclusive  \cite{xu_gapped_2018, zheng_epitaxial_2019}. A larger suite of 
potential  stanene substrates is important to enable robust continued work.

Alumina (Al$_{2}$O$_{3}$) is a wide-gap insulator whose growth is  
well-characterized and commonly performed. Cleaved along its (0001) surface,  
alumina has a surface lattice parameter within a few percent of the 
free-standing stanene lattice parameter. Previous work has examined one 
possible  structure for stanene on Al$_{2}$O$_{3}$ and elucidated basic aspects 
of the  resulting electronic bands \cite{wang_possibility_2016,  
araidai_first-principles_2017}. In this paper, we describe several critical  
results regarding the structure, stability, and topological character of  
stanene on alumina: hexagonal stanene (the assumed structure in the prior  
works) is indeed stabilized on stanene compared to other structures that are  
favored as isolated 2D sheets, the strength of the binding of stanene to  the 
alumina surface turns out to be surprisingly large, the binding is strong  
enough to create an epitaxial 2D layer of stanene on alumina, and the resulting 
 electronic bands of the heterostructure show a large gap as well as the 
desired  topological character of a quantum spin Hall insulator. We end with an 
outlook  for the potential of stanene synthesis on alumina.

\section{Methods}

We performed density functional theory (DFT) calculations
using the Quantum {\sc  Espresso} software package
\cite{giannozzi_quantum_2009}. We used fully
relativistic  projector augmented-wave (PAW) pseudopotentials with spin--orbit 
interaction, along with the Perdew--Burke--Ernzerhof (PBE) 
generalized gradient  approximation to the exchange--correlation functional  
\cite{perdew_generalized_1996}. We used a plane-wave basis set with a
wavefunction  
energy cutoff of 680~eV and a charge density plane wave cutoff of 6,800~eV,
and we relaxed atomic positions until all axial forces were below  
$2.5\times10^{-3}\,\text{eV}/\text{\AA}$. We performed additional calculations 
using the same parameters and a hybrid exchange--correlation functional using the VASP software 
\cite{kresse_efficiency_1996, heyd_hybrid_2003, krukau_influence_2006}.
We performed calculations at the
theoretical relaxed lattice parameters of bulk alumina; however, since the
Quantum {\sc Espresso} version we used does not perform automated
variable-cell relaxations with fully 
relativistic pseudopotentials, those  lattice parameters were found by 
atomically relaxing bulk structures on a grid  of lattice parameter values and 
fitting to find the minimum in energy. We estimate that this is equivalent to 
performing an automated  variable-cell relaxation until all uniaxial stresses 
are below 5~kbar. Calculations used a $12\times12\times1$ $k$-point mesh and 
14~meV  of Gaussian thermal broadening.

We carried out substrate-based calculations on an Al-terminated slab of  
Al$_{2}$O$_{3}$ cleaved along the (0001) surface. In-plane lattice  parameters 
were taken from a theoretical relaxation of bulk Al$_{2}$O$_{3}$,  which 
yielded a lattice parameter of 4.792~\AA. We included four stoichiometric layers
of the Al$_{2}$O$_{3}$ slab to ensure convergence in atomic positions 
and formation energies. We placed monolayers of stanene on both  surfaces of 
a symmetric alumina slab to retain inversion symmetry and avoid  the need for a dipole 
correction in the vacuum. We used the Grimme DFT-D2, DFT-D3, and Becke--Johnson
XDM semiempirical functionals to investigate the robustness of our results against noncovalent interactions between the substrate and the stanene 
overlayer \cite{grimme_semiempirical_2006, barone_role_2009, grimme_consistent_2010, becke_exchange-hole_2007, otero-de-la-roza_van_2012}.

For isolated 2D tin-based monolayers, we computed topological characters from
occupied band parities at time-reversal invariant momenta using the method of 
Fu and Kane \cite{fu_topological_2007}. To compute the topological invariant for
bound stanene, we removed one stanene monolayer from one side of the alumina slab and
used the Wannier
charge center method of Soluyanov and Vanderbilt \cite{soluyanov_computing_2011},
as implemented in the WannierTools package \cite{wu_wanniertools_2018} using
maximally localized Wannier functions from the Wannier90 package 
\cite{pizzi_wannier90_2020}. This approach breaks the inversion symmetry that
was present before, but the resulting electric dipole is quite small and does not
affect the states near the Fermi energy.

\section{Results}

\subsection{Free-standing stanene monolayers}

We performed variable-cell structural relaxations for free-standing monolayers 
of bare stanene, as well as fully functionalized fluorinated stanene (SnF) and 
hydrogenated stanene (SnH). In each calculation, both the lattice parameter and 
the atomic positions were relaxed to minimize stresses and forces. Each 
structure is ``low-buckled,'' with a unit cell containing two 
vertically-displaced Sn atoms. The optimal structural parameters and DFT-PBE band gap, shown in Table \ref{tab:fs}, are in good agreement with previous results 
\cite{xu_large-gap_2013}. According to the Fu--Kane method~\cite{fu_topological_2007}, bare and fluorinated stanene are topological insulators, while hydrogenated stanene is a topologically trivial insulator. The topological 
properties of each of these freestanding materials are examined in greater 
detail in Section III.F below.

\begin{table}
\caption{Structural data for free-standing stanene in its bare, 
hydrogenated, and fluorinated forms.}
\begin{tabular}{lddd}
\hline\hline
& \multicolumn{1}{c}{\;Bare stanene\;}
& \multicolumn{1}{c}{\;Hydrogenated stanene\;}
& \multicolumn{1}{c}{\;Fluorinated stanene\;} \\\hline
Lattice parameter $a$ (\AA)  & 4.68       & 4.72      & 5.02 \\
Sn--Sn buckling $b$ (\AA)    & 0.85       & 0.82      & 0.53 \\
Sn--Sn bond length $d$ (\AA) & 2.83       & 2.85      & 2.95 \\
Band gap (eV)                & 0.069      & 0.214     & 0.295 \\
Topologically nontrivial?       & \multicolumn{1}{c}{YES}
& \multicolumn{1}{c}{NO} & \multicolumn{1}{c}{YES} \\
\hline\hline
\end{tabular}
\label{tab:fs}
\end{table}

\subsection{Bound low-buckled structure}

\begin{figure}
\centering
\includegraphics[width=\textwidth]{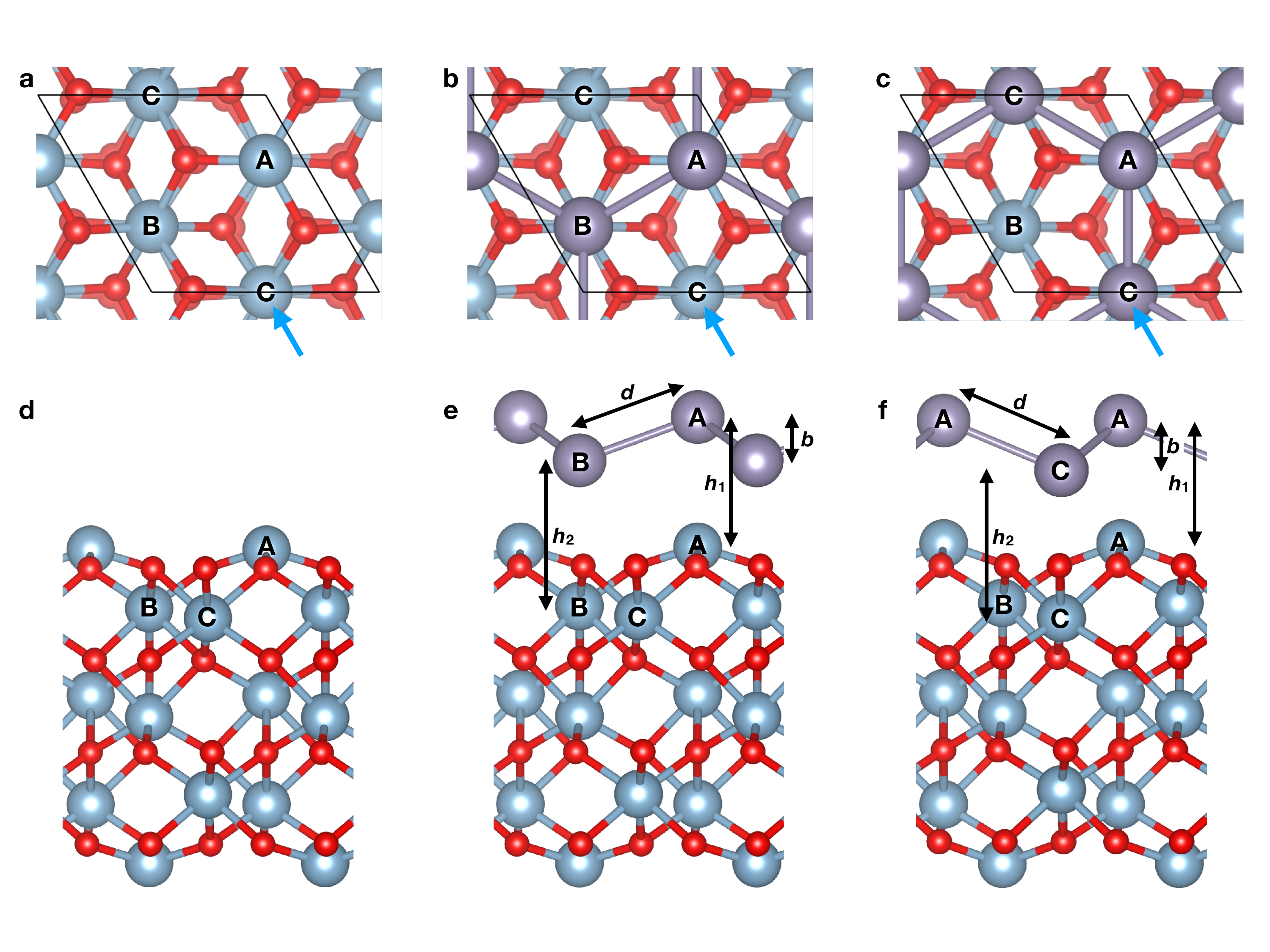}
\caption{\textbf{a,d:} Top (\textbf{a}) and side (\textbf{d}) views of the bare 
Al$_{2}$O$_{3}$(0001) slab used as a substrate. The three inequivalent exposed 
Al atoms are labeled A, B, and C. The blue arrow in (\textbf{a}) indicates the 
viewing direction of panels \textbf{d--f}. \textbf{b,e:} Top (\textbf{b}) and 
side (\textbf{e}) views of second-most-stable registry choice A/B for stanene 
on Al$_{2}$O$_{3}$, placing the upper and lower Sn atoms in positions A and 
B, respectively. Black arrows in \textbf{e} label the bond length $d$, the 
buckling $b$, and the vertical binding distances $h_1$ and $h_2$, whose values 
are found in Table~\ref{tab:struc}. \textbf{c,f:} Top (\textbf{c}) and side 
(\textbf{f}) views of the most stable registry choice A/C for stanene on 
Al$_{2}$O$_{3}$, placing the upper and lower Sn atoms in positions A and C, 
respectively.}
\label{fig1}
\end{figure}

\begin{table}
\caption{Structural and energetic information for the two most favorable 
registries of stanene on Al$_{2}$O$_{3}$. See Fig.~\ref{fig1} for definitions of $b$, $d$, $h_1$ and $h_2$.}
\begin{tabular}{ ldd }
\hline\hline
& \multicolumn{1}{c}{\;A/B structure\;}
& \multicolumn{1}{c}{\;A/C structure\;} \\\hline
Buckling $b$ (\AA)                                   & 1.03  & 1.18  \\
Bond length $d$ (\AA)                                & 2.95  & 3.01  \\
Binding distance $h_1$ (\AA)                         & 3.06  & 2.90  \\
Binding distance $h_2$ (\AA)                         & 3.43  & 3.47  \\
Binding energy $E_b$ per unit cell (eV): \\
\qquad\qquad no van der Waals functional             & 0.31  & 0.50  \\
\qquad\qquad Grimme DFT-D2 functional \cite{grimme_semiempirical_2006}
                                                     & 1.02  & 1.26  \\
\qquad\qquad Grimme DFT-D3 functional \cite{grimme_consistent_2010}
                                                     &       & 0.84  \\
\qquad\qquad XDM functional \cite{becke_exchange-hole_2007, otero-de-la-roza_van_2012}
                                                     &       & 1.16  \\
Band gap (eV)                                        & 0.247 & 0.263 \\
\hline\hline
\end{tabular}
\label{tab:struc}
\end{table}

For our substrate-bound calculations, we focused on undecorated stanene. When 
bound epitaxially to alumina, low-buckled stanene retains its basic structure 
but is  under $\sim$2.4\% tensile strain. We found that the most stable 
structures are obtained when Sn atoms are placed atop Al atoms.

The Al-terminated alumina slab has three exposed aluminum atoms per unit cell, 
which are labeled A, B, and C in Figure~\ref{fig1}(a). Atom A terminates the 
slab, while atoms B and C are roughly coplanar ($\sim$0.2 \AA\ vertical 
separation) and located under a layer of oxygen atoms. We examined a $3\times3$ 
grid of possibly registry alignments for stanene within the alumina unit cell, 
each of which permits two structures that are obtained by swapping the 
up-buckled and down-buckled Sn atoms. We relaxed the atomic positions in each
of these 18 inequivalent stanene-on-alumina registries. The
two most favorable registries, shown in Figures~\ref{fig1}(b) and 
\ref{fig1}(c), place the upper tin atom directly over atom A and the lower tin 
atom directly over either atom B or C. The structural parameters, binding 
energies, and DFT-PBE band gaps of the two favorable structures are found in Table~\ref{tab:struc}. 
The A/C structure is the most energetically favored by a margin of at least 0.24~eV per 
two-atom stanene unit cell. This structure, which was predicted by similar 
previous work \cite{wang_possibility_2016, araidai_first-principles_2017}, will 
be taken as the ground state structure.

\subsection{Binding energy and van der Waals functionals}

The binding energy $E_b$ equals the total energy of the bound 
stanene-substrate complex $E_\text{bound}$, minus the sum of the energies of 
the free-standing stanene layer $E_{\text{stanene}}$ and the bare alumina slab 
$E_{\text{Al}_2\text{O}_3}$:
\begin{equation}
E_b=E_\text{bound}-\left(E_\text{stanene}+E_{\text{Al}_2\text{O}_3}\right).
\end{equation}
To assess the importance of noncovalent interactions in the binding, we
calculated $E_b$ both with and without van der Waals dispersion corrections.
We checked three van der Waals functionals implemented in Quantum {\sc Espresso}:
the common Grimme DFT-D2 functional \cite{grimme_semiempirical_2006};
its DFT-D3 revision, which incorporates three-body interactions \cite{grimme_consistent_2010};
and the Becke--Johnson exchange-hole dipole-moment (XDM) model
\cite{becke_exchange-hole_2007, otero-de-la-roza_van_2012}.
We found that including a van der Waals functional modifies the interatomic
distances listed in Table~\ref{tab:struc} 
by less than 0.5\%, indicating that the physical structure is determined 
largely by chemical rather than van der Waals interactions. However, the 
binding energy, which is 0.50~eV per stanene unit cell without dispersion effects,
increase to 0.84--1.23~eV per unit cell depending on the dispersion functional
used. The largest binding occurs with the DFT-D2 functional which is known
to overbind solids \cite{grimme_consistent_2010}, so we expect that the true binding
energy lies within the range between the bare and DFT-D2 calculations.
A previous study using
the optB86b-vdW functional found a binding energy of 1.11~eV per unit cell which is well within this range \cite{wang_possibility_2016}.
This suggests that both 
noncovalent and covalent interactions are needed to fully describe the absolute 
magnitude of the binding energy of stanene to alumina.

\begin{figure}
\centering
\includegraphics[width=\textwidth]{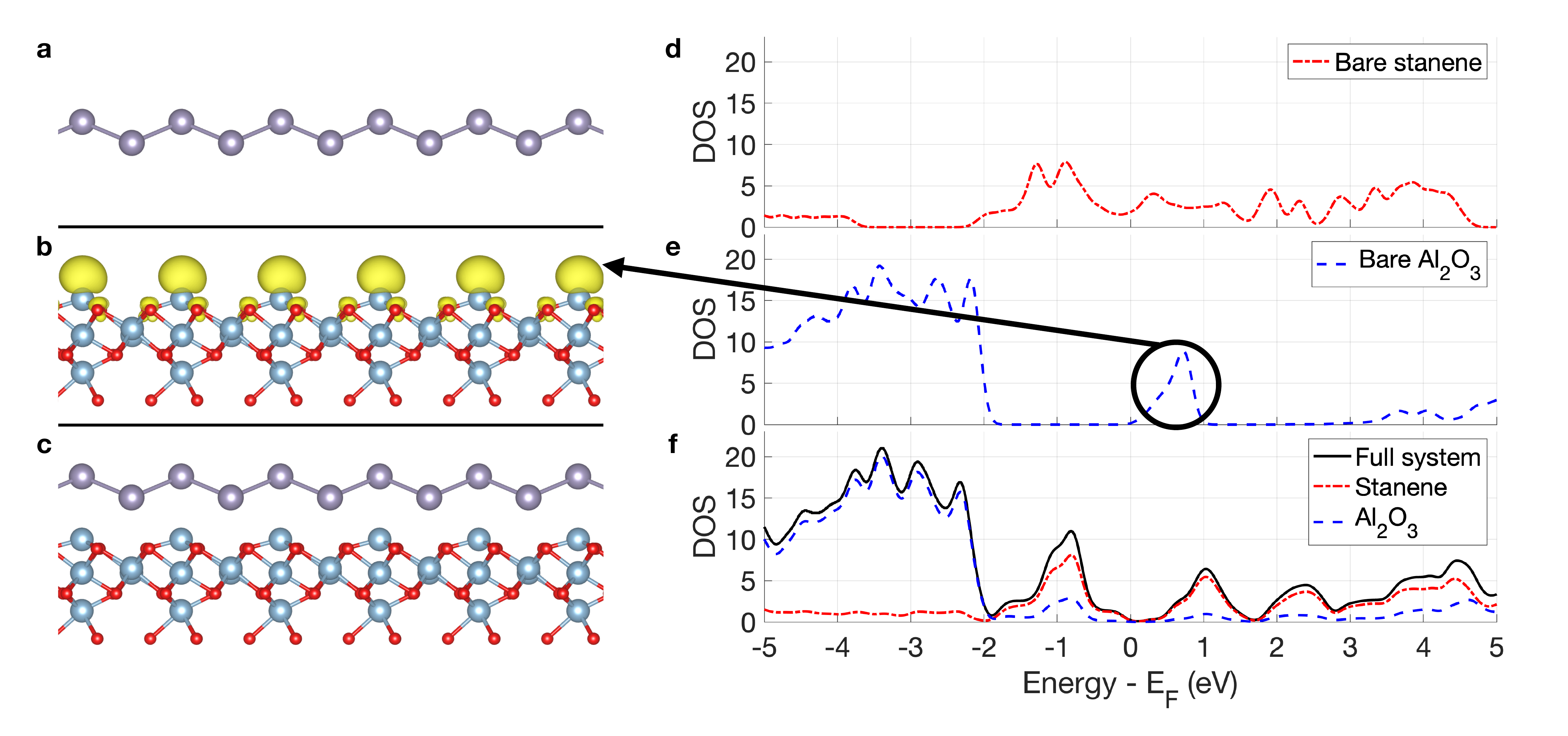}
\caption{\textbf{a,c:} Side view of free-standing stanene layer and stanene 
bound to the Al$_{2}$O$_{3}$(0001) slab, respectively. \textbf{b:} Side view of 
bare alumina slab. The yellow features are an isosurface of the local density 
of states (LDOS) integrated from the Fermi level ($E_F$) to 1.0 eV above $E_F$. 
\textbf{d,e,f:} Density of states (DOS) plots for free-standing stanene, bare 
alumina, and bound stanene on alumina, respectively. In the plot for the full 
system, the DOS is also projected onto the L\"owdin orbitals of the tin atoms (stanene)
and the aluminum/oxygen atoms (Al$_2$O$_3)$.}
\label{fig2}
\end{figure}

The fact that the chemical binding is quite substantial at 0.50 eV/unit cell 
requires some explanation: naively, one might expect a wide-gap material such 
as alumina to be relatively inert. To identify the chemical interaction that 
drives the binding, we plotted the density of states (DOS) of the free-standing 
stanene layer, the bare alumina slab, and the stanene-substrate complex 
(Figures~\ref{fig2}(d-f)). The bare alumina slab displays a peak in the DOS 
just above the Fermi level, which represents a surface state localized to 
``dangling'' orbitals on the top layer of exposed Al atoms (left panel of 
Figure~\ref{fig2}(b)). This state vanishes upon the binding of stanene---as can 
be seen from  Figure~\ref{fig2}(f), the states of the full complex near the 
Fermi level are dominated by Sn orbitals. The unoccupied alumina orbital 
hybridizes with various Sn orbitals, spreading out in energy over the former 
alumina gap. In particular, a portion of this orbital forms a new bonding 
orbital between $-2$ and 0 eV in Figure~\ref{fig2}(f).

\begin{figure}
\centering
\includegraphics[width=\textwidth]{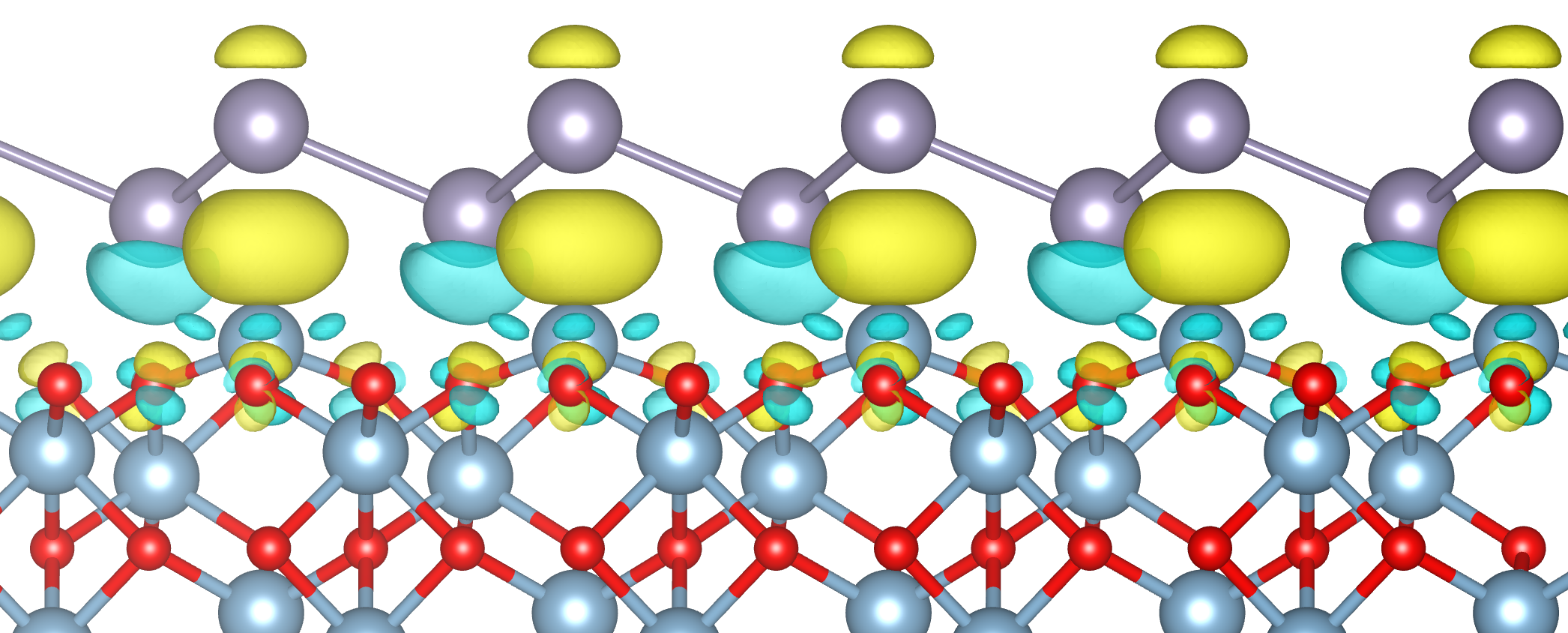}
\caption{Side view of stanene on alumina showing positive (yellow) and negative 
(cyan) isosurfaces for electron redistribution. For regions in yellow, the 
electron density of stanene on alumina is greater than the sum of the electron 
density of the bare slab and that of the free-standing stanene monolayer, 
indicating an increase in electron density during binding. For regions in cyan, 
the reverse is true.}
\label{fig3}
\end{figure}

We confirmed that the originally empty ``dangling'' states of the alumina slab 
remain localized to the vicinity of the exposed Al atom by examining the 
redistribution of electron density shown in Figure~\ref{fig3}. During binding, 
electron density redistributes from the cyan regions to the yellow regions, 
including a large $\sigma$-like region between the surface Al atom and the Sn 
atom above it. This indicates that the formerly-vacant Al orbital becomes 
filled as it moves lower in energy and hybridizes with nearby Sn orbitals: a 
heteropolar covalent bond has formed, explaining the substantial binding energy 
of 0.50 eV even without van der Waals interactions.

\subsection{Commensurate versus incommensurate binding}

Since stanene must be placed under 2.4\% tensile strain to bind epitaxially to 
the Al$_{2}$O$_{3}$(0001) surface, we checked for the possibility of 
incommensurate rather than epitaxial binding. In a case of incommensurate 
binding at the free-standing lattice parameter, the Sn monolayer will be 
unstrained, but most of the monolayer's area will not attain its preferred 
registry with respect to the alumina substrate. Therefore, the competition 
between epitaxial and incommensurate binding depends on a comparison of the 
strain energy of the stanene monolayer to the energy penalty for placing the 
monolayer on a non-optimal registry.

The energy to strain a free-standing stanene monolayer from its equilibrium  
lattice parameter of $a=4.68\,\text{\AA}$ to the theoretical Al$_{2}$O$_{3}$  
lattice parameter of $a=4.792\,\text{\AA}$ is 25~meV per unit cell. For low 
strain, an incommensurate overlayer on a substrate can be treated as  a 
long-wavelength superlattice, with each unit cell of the superlattice  sampling 
a different registry. If we label the in-plane position of the lower  Sn atom 
(Sn atoms B or C in Figure~\ref{fig1}) by  
$\mathbf{x}=u_1\mathbf{a}_1+u_2\mathbf{a}_2$, then the average energy of a  
single unit cell in such an incommensurate overlayer is approximately
\begin{equation}
E_\text{incomm}
=\int_0^1\int_0^1E\left(u_1\mathbf{a}_1+u_2\mathbf{a}_2\right)du_1\,du_2,
\end{equation}
where $E(\mathbf{x})$ is the energy of a single commensurate unit cell with a 
lower Sn atom placed at $\mathbf{x}$. We estimate this integral using the 
registry calculations performed earlier, considering only the $3\times3$ grid 
of structures that are lateral shifts of the optimal A/C structure.
The incommensurate energy penalty is 333 meV per unit cell with no van der
Waals functional employed, 458 meV per unit cell with the Grimme DFT-D2
functional, and 339 meV per unit cell with the Grimme DFT-D3 functional.
In all cases, the incommensurate energy is an order of magnitude greater than
the strain energy, so we conclude that the incommensurate structure is
irrelevant and that stanene will bind epitaxially on the alumina substrate.

\subsection{Dumbbell stanene}

We considered the ``dumbbell'' stanene structure proposed by Tang et al., which 
contains 10 Sn atoms in a multilayered analogue of a $2\times2$ stanene 
supercell \cite{tang_stable_2014}. In an isolated monolayer, out-of-plane 
$sp^3$ hybridization renders dumbbell stanene lower in energy than low-buckled 
stanene by 0.18~eV per Sn atom \cite{tang_stable_2014}. However, we find that 
dumbbell stanene binds only weakly to Al$_{2}$O$_{3}$, collapsing into a 
disorderly structure that is 0.46~eV per Sn atom higher in energy than the 
bound low-buckled configuration. This occurs because the highly buckled 
dumbbell structure prevents a close wetting interaction between Sn atoms and 
the substrate. Therefore, dumbbell stanene is not a relevant 
phase when considering epitaxial stanene on alumina.

\subsection{Free-standing and bound stanene band structures}

\begin{figure}
\centering
\includegraphics[width=\textwidth]{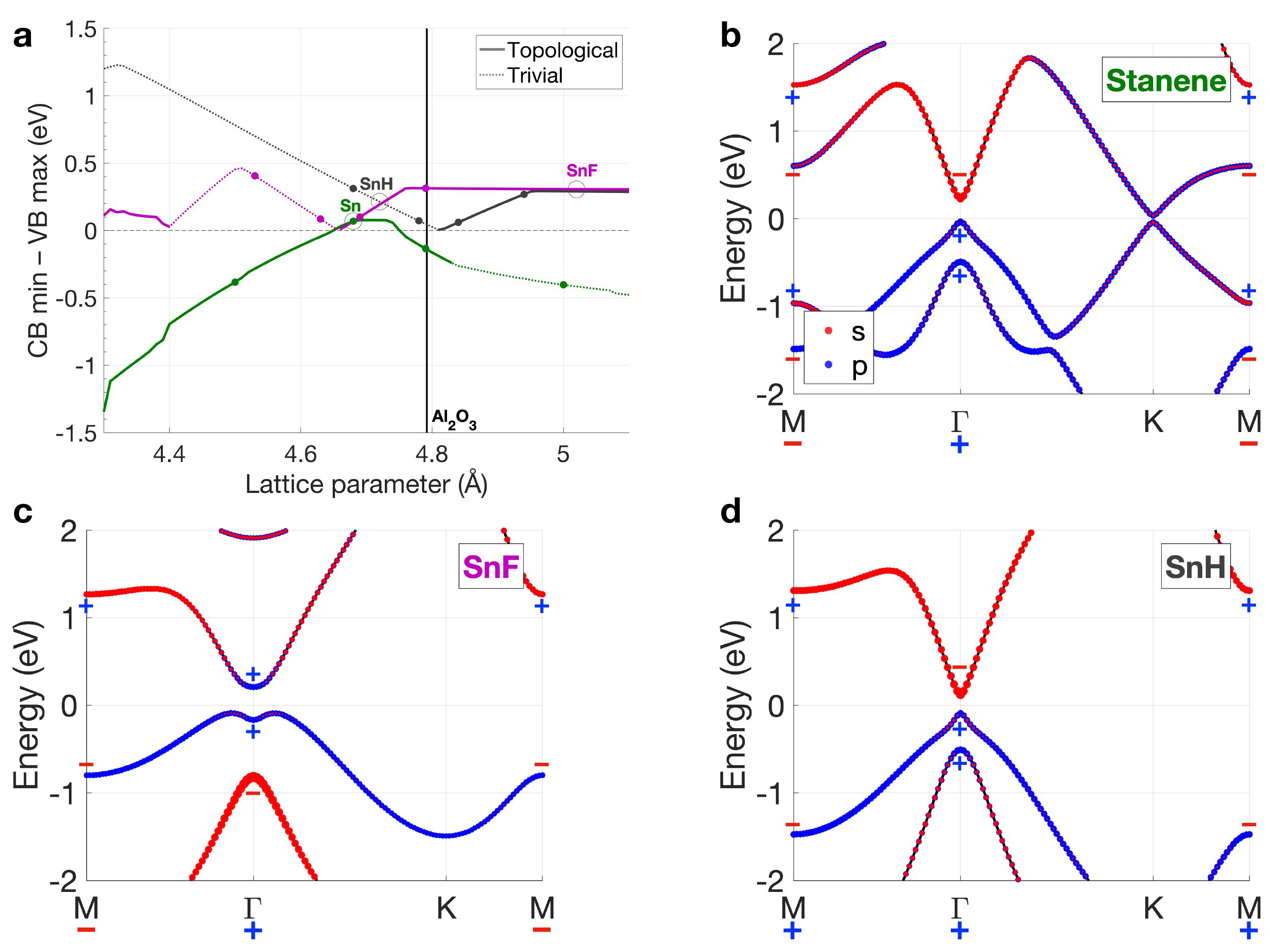}
\caption{\textbf{a:} Evolution of the band gap of bare stanene (green), 
fluorinated stanene SnF (magenta), and hydrogenated stanene SnH (black) as a 
function of lattice parameter. The band gap is defined as the signed energy
difference between the conduction band (CB) minimum and the valence band (VB)
maximum, meaning that it is negative for semimetallic materials. The equilibrium
lattice parameter of Al$_{2}$O$_{3}$ is indicated with a vertical black line. 
For each structure, a dashed line indicates a trivial material with a topological
index of $Z_2=0$, while a solid line indicates a nontrivial material with
$Z_2=1$. Each material's equilibrium lattice parameter is marked with a large open
circle, while the band structures plotted in Figure~\ref{fig5} correspond to the
points marked with small filled circles. \textbf{b,c,d:} Band 
structures of bare stanene, fluorinated stanene SnF, and hydrogenated stanene 
SnH at their equilibrium lattice parameters. Bands are colored by 
their $s$-orbital (red) and $p$-orbital (blue) characters, and labeled with their
parities at time-reveral invariant momenta (TRIMs). On the $x$-axis, each TRIM is
labeled with the product of all of its band parities, the quantity called
$\delta_i$ in the Fu--Kane treatment \cite{fu_topological_2007}. The zero of band
energy is the Fermi level $E_F$.}
\label{fig4}
\end{figure}

Next, we performed a thorough investigation of the band structures and 
topological indices of free-standing stanene and its derivatives. Figure~\ref{fig4}(a) shows the 
evolution of the band gaps of free-standing bare stanene, fluorinated stanene, 
and hydrogenated stanene as a function of lattice parameter, highlighting 
regimes in which each material is a topological insulator, a topologically nontrivial material with negative gap (defined below), and a topologically trivial metal. For metallic stanene layers, 
the band structures exhibit a semimetallic negative gap since the 
conduction band minimum drops below the valence band maximum compared to nearby 
insulating structures, and for them we compute the topological index that we 
would obtain if the valence and conduction bands were pulled apart far enough 
to create a global gap without further modification of the electronic 
structure. Figures~\ref{fig4}(b-d) show the equilibrium band structures of
free-standing stanene, fluorinated stanene SnF, and hydrogenated stanene SnH (also called stanane \cite{xu_large-gap_2013}).

\begin{figure}
\centering
\includegraphics[width=\textwidth]{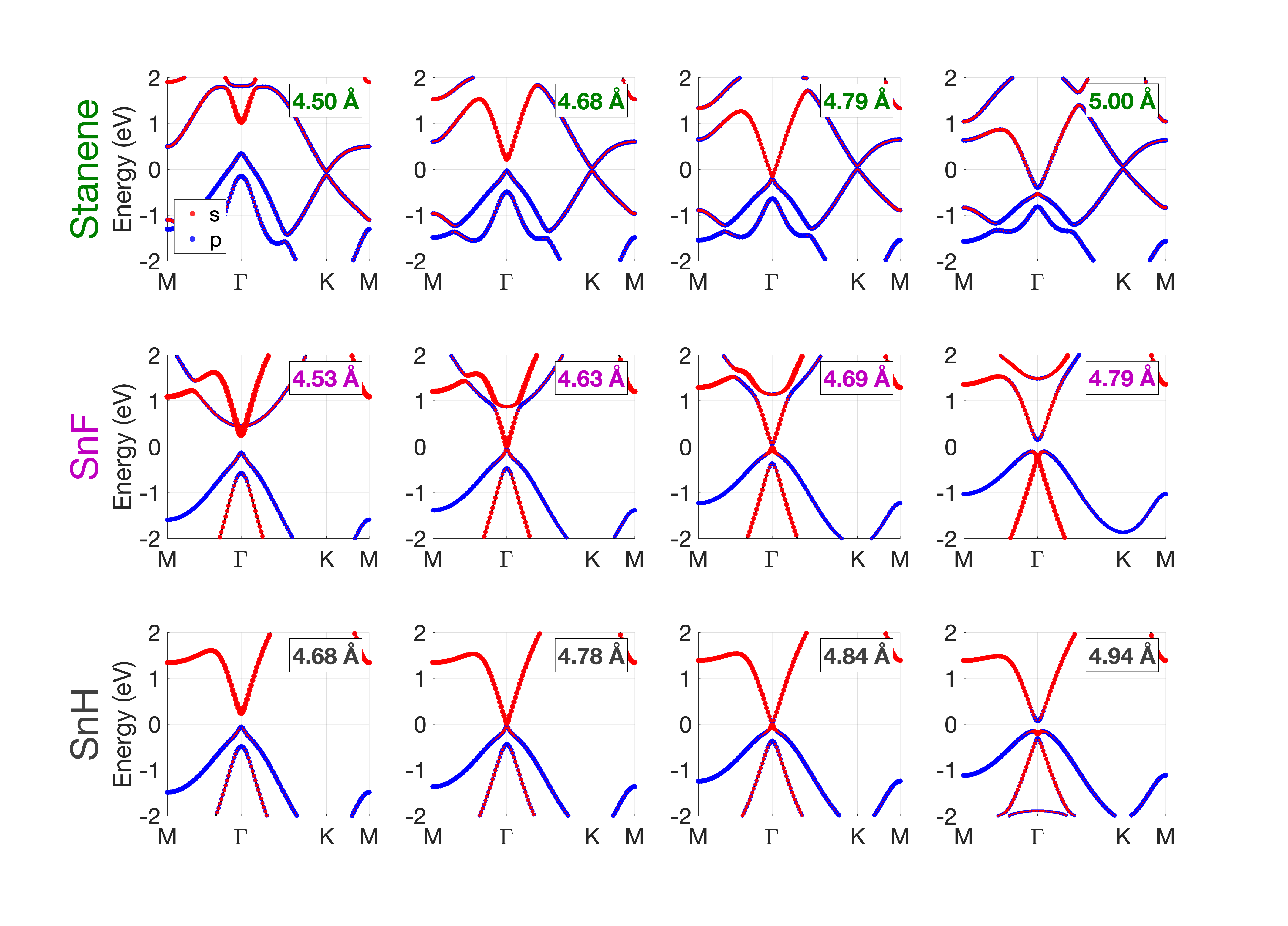}
\caption{Band structure plots showing the change of topological character via 
band inversions in bare stanene (top row), fluorinated stanene SnF (middle row),
and hydrogenated stanene SnH (bottom row). The zero of energy in each plot is the
Fermi level $E_F$. In each 
case, a negative-parity $s$-band crosses down from the conduction band into 
the valence band as the lattice parameter increases. Due to their different 
equilibrium lattice parameters, free-standing stanene ($a=4.68\,\text{AA}$) and SnF
($a=5.02\,\text{\AA}$) are topological insulators, while SnH ($a=4.72\,\text{\AA}$)
is a trivial insulator, but each material can be tuned to the other regime using strain.}
\label{fig5}
\end{figure}

The band structure plots in Figure~\ref{fig5} illustrate how the bands of each material
evolve under strain. In the case of bare stanene (top row), the valence band at $\Gamma$
sits well above the Dirac cone at K when compressive strain is applied ($a=4.50\,\text{\AA}$),
resulting in a negative-gap semimetal. As the lattice parameter increases ($a=4.68\,\text{\AA}$),
the gaps at $\Gamma$ and K line
up, forming a globally gapped topological insulator. For small tensile strain 
($a=4.79\,\text{\AA}$), the gap at $\Gamma$ is pushed below the Dirac cone at K: 
this is the regime of strain relevant for stanene on Al$_{2}$O$_{3}$, so the 
substrate is necessary to open the global band gap (we will see that is also  
sufficient below). Finally, for tensile strain somewhat larger than that 
applied by Al$_{2}$O$_{3}$ ($a=5.00\,\text{\AA}$), the gap closes between the 
negative-parity $s$-type conduction band at $\Gamma$ and the positive-parity 
$p$-type valence band (\emph{i.e.}, a band inversion occurs at $\Gamma$). The resulting 
parity exchange renders free-standing stanene topologically trivial above a lattice 
parameter of roughly 4.83~\text{\AA}.

The story is slightly different for SnF and SnH. At their equilibrium lattice parameters, 
SnF is a topological insulator while SnH is a topologically trivial insulator. However, the 
two materials are actually quite similar electronically: both materials are 
trivial insulators under sufficient compressive strain and topological 
insulators under sufficient tensile strain. The difference between them at 
equilibrium is simply due to the relative ordering of the lattice parameter of 
the topological transition and the equilibrium lattice parameter. This can be seen
clearly in the middle-row and bottom-row band structures of Figure~\ref{fig5} which
depict the topological phase transitions in the two materials. In each case, a
negative-parity antibonding band constructed from Sn $s$ orbitals moves down through the 
conduction band, and crosses over to the valence band at $\Gamma$, inducing a 
band inversion and leading to a nontrivial topological index.

Since each topological phase transition is controlled by a band inversion across
the gap at $\Gamma$, it is reasonable to ask how robust our results are against a method that accounts better for electron--electron interactions. To check this, we calculated
the band structure of relaxed, topologically nontrivial stanene and SnF with VASP \cite{kresse_efficiency_1996}
using the HSE06 hybrid functional \cite{heyd_hybrid_2003, krukau_influence_2006}, with detailed band structures available in the Appendix (Figure~\ref{fig8}). The band gaps in both materials (Figures~\ref{fig4}(b,c)) arise from spin--orbit coupling between $p$-states at K and $\Gamma$, respectively, so the incorporation of the HSE06 does not change the qualitative shape of the immediately $E_F$-adjacent bands. The primary qualitative change is that the HSE06 function shifts $s$-states up and $p$-states down at $\Gamma$ (Table~\ref{tab:hybrid}). In spite of this change, the proper band inversions are preserved and both materials remain topologically nontrivial under the HSE06 functional.

\begin{figure}
\centering
\includegraphics[width=\textwidth]{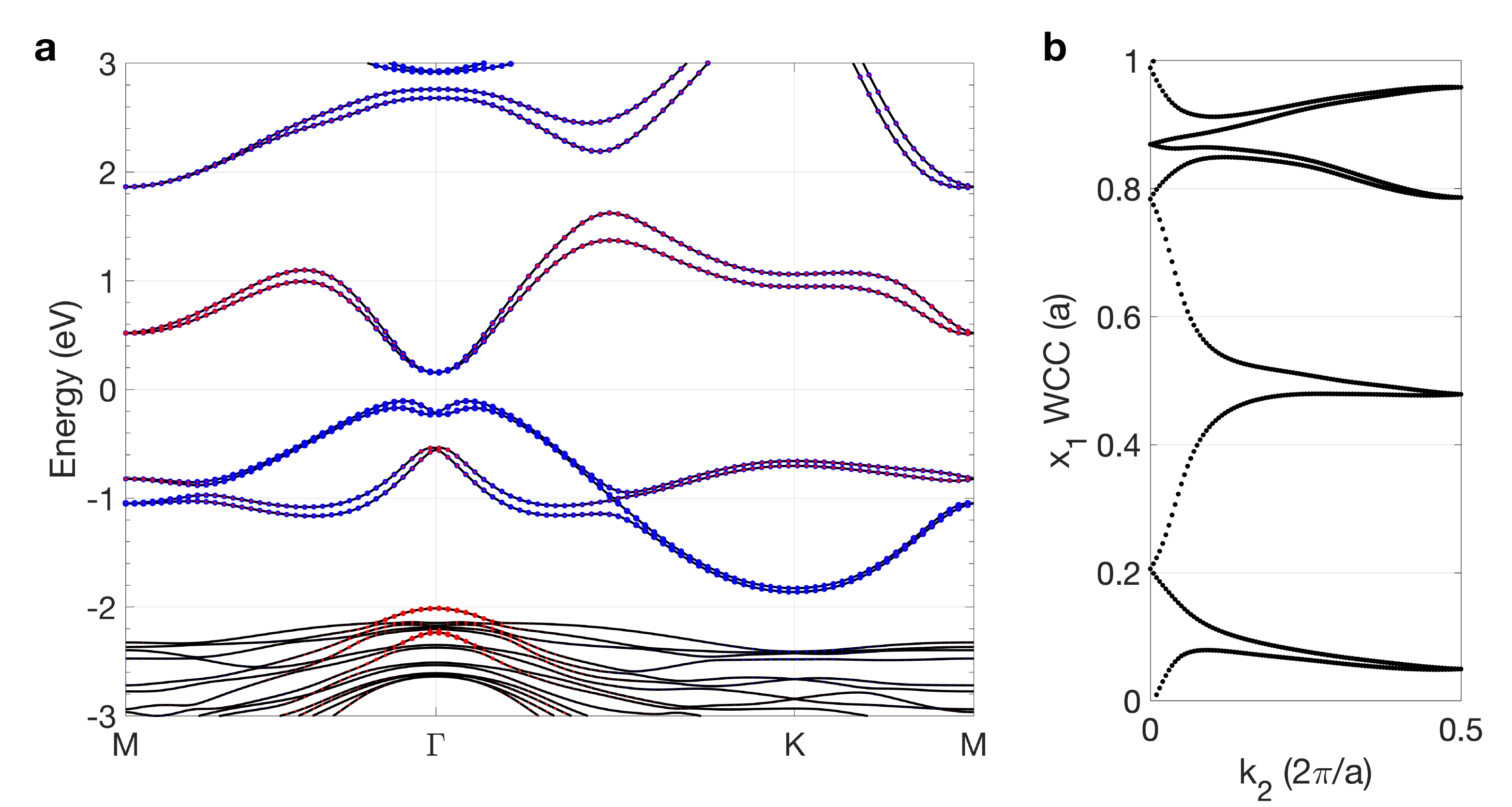}
\caption{Calculations of epitaxial stanene on alumina. \textbf{a:} DFT band structure plot: bands are colored by their Sn $s$-orbital (red) and Sn $p$-orbital (blue) characters. The DFT band gap is calculated to be 0.263~eV. \textbf{b:} Evolution of the Wannier charge centers (WCCs) in the $x$-$y$ plane, computed according to the technique in \cite{soluyanov_computing_2011}. The winding of the WCCs demonstrates that bound stanene is a topological insulator.}
\label{fig6}
\end{figure}

Figure~\ref{fig6}(a) shows the the band structure of the full stanene-on-alumina 
system. It differs from the bare-stanene band structure (Figure~\ref{fig4}(b)) 
in several important ways. First, the presence of the substrate breaks 
inversion symmetry, which, when combined with the spin-orbit interaction, 
leads to a Rashba splitting of the conduction bands away from the $\Gamma$ 
point. We computed the  band structures for stanene on alumina slabs of different 
thickness and confirmed that the $k$-dependent energy splitting of bands away 
from the $\Gamma$ point is due to inversion symmetry breaking (rather than 
evanescent coupling between the two surfaces of the alumina slab). This data
is available in the Appendix.

Finally, the Dirac cone at K has vanished due to the partial saturation of the 
stanene $p_z$ orbitals by the alumina surface, and the ordering of $s$-type and 
$p$-type orbitals at $\Gamma$ has become inverted, yielding a band structure
remarkably similar to the free-standing SnF band structure. These features
are necessary for the existence of topological behavior in stanene on a substrate 
\cite{xu_large-gap_2013}. Indeed, we confirmed that stanene on alumina is
topologically nontrivial using the Wannier charge center method (Figure~\ref{fig6}(b)) \cite{soluyanov_computing_2011}.

\subsection{\emph{In situ} stanene fluorination}

\begin{figure}
\centering
\includegraphics[width=\textwidth]{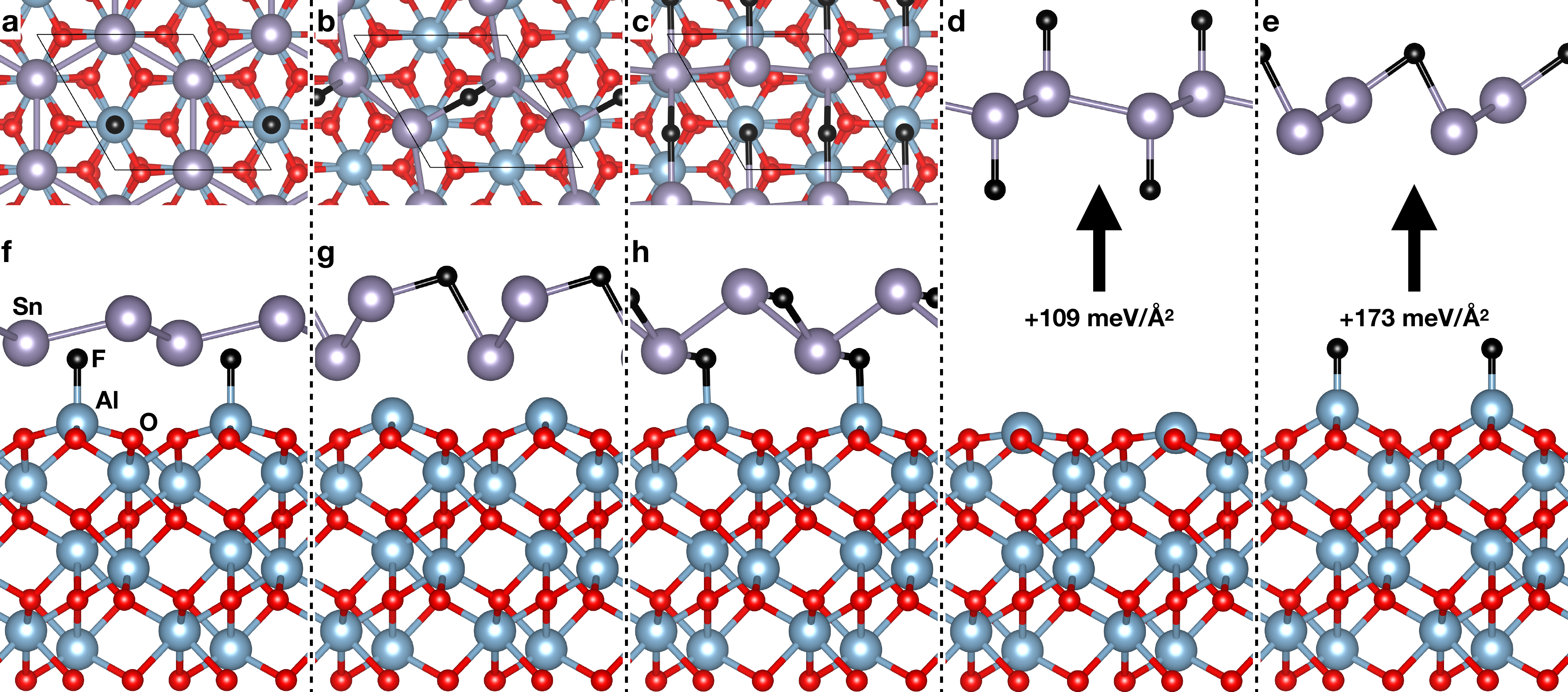}
\caption{\emph{In situ} fluorination of the stanene sheet. \textbf{a,f:} Top (\textbf{a}) and side (\text{f}) views of the half-fluorinated stanene sheet. \textbf{b,g:} Top (\textbf{b}) and side (\textbf{g}) views of the higher-energy stanene sheet fluorinated at the bridging site but not at the Al-bound site. \textbf{c,h:} Top (\textbf{c}) and side (\textbf{h}) views of the fully-fluorinated stanene sheet in the Al$_{2}$O$_{3}$-bound ground state. \textbf{d:} Illustration of clean SnF exfoliation. \textbf{e:} Illustration of exfoliation of the half-fluorinated stanene sheet, with the bound F atom left behind. The energy cost for this process is higher than for clean exfoliation.}
\label{fig7}
\end{figure}

In our final set of calculations, we consider the possibility of \emph{in situ} functionalization of stanene directly on the Al$_{2}$O$_{3}$ surface as a route for future experimental fabrication. As discussed in Section III.F, the Al$_{2}$O$_{3}$ substrate and decorating atoms (e.g., F) interact chemically with the Sn hexagonal lattice in analogous ways. Therefore, we expect a fluorinated tin monolayer to take on some structure other than the pristine free-standing SnF sheet on a substrate. However, the pristine structure of isolated SnF should appear if the monolayer is removed from the Al$_{2}$O$_{3}$ surface or deposited on a more inert substrate, as it is the ground state of a two-dimensional crystal of stoichiometrically balanced Sn and F. Therefore, a serial process of fluorination followed by exfoliation could, in principle, yield topologically nontrivial SnF via deposition on Al$_{2}$O$_{3}$.

We obtained the ground-state structures for stanene on Al$_{2}$O$_{3}$ under fluorine half-functionalization (one F per surface unit cell, Figures~\ref{fig7}(a,f)) and full-functionalization (two F per surface unit cell, Figures~\ref{fig7}(c,h)). In the fully-functionalized SnF ground state, one F atom sits over the exposed Al atom (site A in the notation of Figure~\ref{fig1}), while the other F atom occupies a ``bridging'' position between the Sn atoms, which remain buckled. The half-fluorinated ground state places the F atom at the exposed Al site, so we infer that this site will be occupied first during fluorination, with the bridging site being filled later. Accommodating the position of the Al-bound F atom requires a registry shift of the stanene sheet (compare Figures~\ref{fig1}(c,f)) that, without F present, costs 41~meV/\AA$^{2}$ in energy. The half-fluorinated structure with the F in the bridging position (Figures~\ref{fig7}(b,g)) is also an energetic local minimum, but is 0.91~eV per unit cell higher in energy than the Al-bound structure.

Bound, fully fluorinated SnF binds to the alumina surface with a binding energy of 109~meV/\AA$^{2}$, substantially more than the 63~meV/\AA$^{2}$ for bare stanene (see Table~\ref{tab:struc}). For layered bulk materials, both of these binding values lie in the range considered ``potentially exfoliable'' (Figure~\ref{fig7}(d))\cite{mounet_two-dimensional_2018}. One worry we may entertain is that after full-functionalization, the Al-bound F atom could remain in place during exfoliation, yielding a singly-fluorinated Sn sheet (Figure~\ref{fig7}(e)). However, from an energetic perspective, this is unlikely: such a structure is 173~meV/\AA$^{2}$ higher in energy than the fully decorated, bound structure, taking it well out of the range of potential exfoliability. Therefore, exfoliation post functionalization with F is expected to produce a fully-fluorinated stanene sheet. Experimental realization of this exfoliation represents an interesting future endeavor.

\section{Discussion and outlook}

The combination of strong epitaxial binding with band inversion at $\Gamma$ and 
$p_z$-orbital saturation at K indicates that alumina is a promising substrate 
for the synthesis of bare monolayer stanene. Such a material offers an 
opportunity for experimental observation of the quantum spin Hall effect 
\cite{xu_large-gap_2013, zhang_room_2017} as well as a substrate for a variety 
of technological applications \cite{zheng_spin-current_2018, 
marin_tunnel-field-effect_2018, vandenberghe_imperfect_2017, 
abbasi_modulation_2019, abbasi_theoretical_2019}. In addition, the spin 
separation in the conduction band due to the Rasbha splitting can be 
harnessed for applications in spintronics and topological superconductivity 
\cite{wang_two-dimensional_2014}.

Controlled functionalization, e.g., by hydrogen or fluorine, is also an 
important avenue of stanene research, since functionalization both enhances the 
band gap and protects against unwanted environmental interactions with Sn $p_z$ 
orbitals \cite{xu_large-gap_2013}. Generally speaking, two-dimensional 
materials can be synthesized either by epitaxial deposition or by the 
exfoliation of multilayered van der Waals materials \cite{molle_buckled_2017}. 
The latter method is attractive since it is flexible and modular, but is 
impractical for materials like bare stanene whose 3D bulk phase ($\alpha$-tin) 
is not intrinsically layered. As we showed in Section III.G, introducing fluorine to a bound stanene monolayer is expected to form a functonalized tin sheet that may be exfoliable for use in heterostructure and device applications.

In summary, we have shown that monolayer stanene binds strongly and epitaxially 
to the Al$_{2}$O$_{3}$(0001) surface, with a buckled structure and a sizable 
global band gap. We have examined the chemical character of the binding and 
verified the topologically nontrivial nature of stanene on alumina. With its 
wide surface band gap and relative inertness, alumina is a promising substrate 
for future experimental fabrication and characterization.

\begin{acknowledgments}

This work was supported by Function Accelerated nanoMaterial Engineering  
(FAME), AFOSR, No. FA9550-15-1-0472. It also used the Extreme  Science and 
Engineering Discovery Environment (XSEDE), which is supported by  National 
Science Foundation grant number ACI-1548562, via computer time on the  Comet 
supercomputer as enabled by XSEDE allocation MCA08X007. S.E. also  acknowledges 
support from the NSF Graduate Research Fellowship, No.\  DGE1752134. Atomic 
structural images were produced with the VESTA package \cite{momma_vesta_2011}.

\end{acknowledgments}

\appendix*
\section{Additional calculations}

\begin{figure}
\centering
\includegraphics[width=\textwidth]{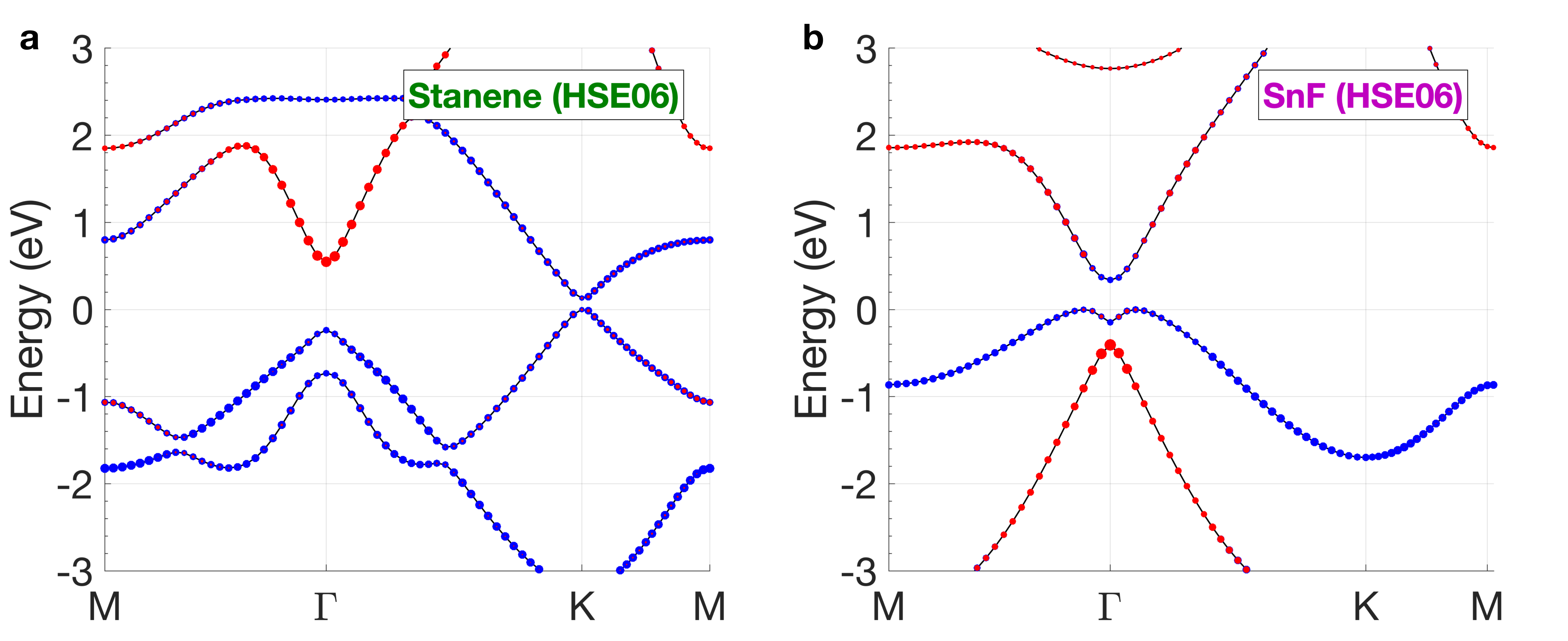}
\caption{Orbital-projected hybrid (HSE06) band structures of stanene and SnF; the analogous non-hybrid calculations are shown in Figures~\ref{fig4}(b-c). The zero of energy is the Fermi level $E_F$.}
\label{fig8}
\end{figure}

Figure~\ref{fig8} shows the band structures of bare and fluorinated stanene
computed using the HSE06 hybrid functional. Bands are labeled by their
$s$ (red) and $p$ (blue) character to demonstrate that the band ordering
that ensures the topological character of each material is preserved, even
when better accounting for nonlocal electron-electron interactions.

\begin{figure}
\centering
\includegraphics[width=\textwidth]{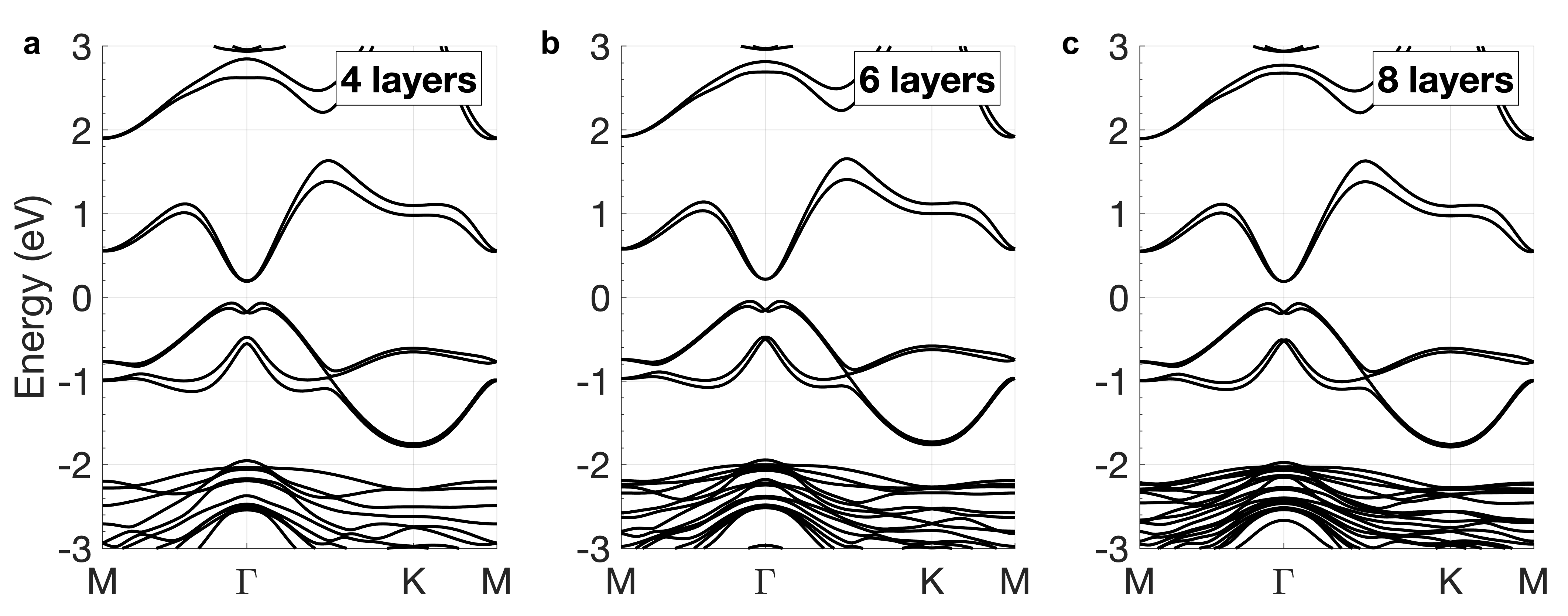}
\caption{Substrate thickness dependence of the stanene-on-alumina band structure.}
\label{fig9}
\end{figure}

Figure~\ref{fig9} compares the band structures of stanene on 4-, 6-, and 8-layer
alumina slabs. Below around $-2$~eV, an increasing density of bulk bands can be seen,
but the near-$E_F$ states, localized to the tin overlayer, are unaffacted
by the thickness of the substrate. 

\begin{table}
\caption{Key band energy differences (eV) for bare stanene and SnF using the PBE and the HSE06 exchange-correlation functionals. ``Valence $p$--$p$ splitting'' is the energy difference between the bands colored blue in Figures~\ref{fig4}(b-c) and \ref{fig8}. ``Valence $s$--$p$ difference'' is the signed energy difference between the $s$-band (red) and the \emph{average} of the two $p$-band (blue) energies.}
\begin{tabular}{rdd|rdd}
\hline\hline
\multicolumn{1}{c}{Stanene} &
\multicolumn{1}{c}{PBE} & \multicolumn{1}{c|}{HSE06} &
\multicolumn{1}{c}{SnF} &
\multicolumn{1}{c}{PBE} & \multicolumn{1}{c}{HSE06} \\\hline
Valence $p$--$p$ splitting @ $\Gamma$ & 0.45 & 0.49 &
Valence $p$--$p$ splitting @ $\Gamma$ & 0.37 & 0.49 \\
Valence $s$--$p$ difference @ $\Gamma$ & +0.48 & +1.04 &
Valence $s$--$p$ difference @ $\Gamma$ & -0.85 & -0.51 \\
Direct band gap @ K & 0.07 & 0.13 \\
\hline\hline
\end{tabular}
\label{tab:hybrid}
\end{table}

\end{document}